\documentclass[twocolumn,twoside,preprintnumbers,amsmath,amssymb,showkeys]{revtex4}

\usepackage{epsfig}

\usepackage{graphicx}

\usepackage{fancyhdr}

\usepackage{pslatex}

\usepackage{algorithm}
\usepackage{algorithmic}

\begin{document}

\title{Electron and Positron Capture Rates on $\bf{^{55}}$Co in Stellar Matter}

\author{Jameel-Un Nabi\footnote{Corresponding author e-mail: jnabi00@gmail.com},
Muneeb-Ur Rahman, and  Muhammad Sajjad}

\affiliation{Faculty of Engineering Sciences,
Ghulam Ishaq Khan Institute, of Engineering Sciences
and Technology, Topi 23640, Swabi, NWFP, Pakistan}

\received{on 22 September, 2007}

\begin{abstract}
$^{55}Co$ is not only present in abundance in presupernova phase but
is also advocated to play a decisive role in the core collapse of
massive stars. The spectroscopy of electron capture and emitted
neutrinos yields useful information on the physical conditions and
stellar core composition. B(GT) values to low-lying states are
calculated microscopically using the pn-QRPA theory. Our rates are
enhanced compared to the reported shell model rates. The enhancement
is attributed partly to the liberty of selecting a huge model space,
allowing consideration of many more excited states in our rate
calculations. Unlike previous calculations the so-called Brink's
hypothesis is not assumed leading to a more realistic estimate of
the rates. The electron and positron capture rates are calculated
over a wide temperature $(0.01\times10^{9} -
 30\times10^{9}K)$ and density $(10 - 10^{11} gcm^{-3})$ grid.


\end{abstract}

\maketitle

\thispagestyle{fancy}
\setcounter{page}{1}

\section{ Introduction}

Weak interactions and gravity decide the fate of a star. These two
processes play a vital role in the evolution of stars. Weak
interactions deleptonize the core of massive star, determine the
final electron fraction $(Y_{e})$, and the size of the homologous
core. The collapse is very sensitive to the entropy and to the
number of leptons per baryons, $Y_{e}$, [1]. Electron capture and
photodisintegration processes in the stellar interior cost the core
energy by reducing the electron density and as a result the collapse
of stellar core is accelerated under its own ferocious gravity.
This collapse of the stellar core is very sensitive to the core entropy
and to the number of lepton to baryon ratio. These two quantities
are mainly determined by weak interaction processes. The simulation
of the core collapse is very much dependent on the electron capture
on heavy nuclides [2]. When the stellar core attains densities close
to $10^{9}$ $gcm^{-3}$, it consists of heavy nuclei imbued in
electrically neutral plasma of electrons, with small fraction of
drip neutrons and an even smaller fraction of drip protons [3]. At
this stage the density of the stellar core is much lower than the
nuclear matter density and thus the average volume available to a
single nucleus is much greater than that of nuclear volume. Electron
capture and beta decay decide the ultimate fate of the star. During
the stellar core collapse, the entropy of the stellar core decides
whether the electron capture occur on heavy nuclei or on free
protons produced in the photodisintegration process. Stars with mass
$>$ $8M_\odot$ after passing through all hydrostatic burning stages
develop an onion like structure and produce a collapsing core at the
end of their evolution and lead to increased nuclear densities in
the stellar core [4]. Electron capture on nuclei takes place in very
dense environment of the stellar core where the Fermi energy
(chemical potential) of the degenerate electron gas is sufficiently
large to overcome the threshold energy given by negative Q values of
the reactions involved in the interior of the stars. This high Fermi
energy of the degenerate electron gas leads to enormous electron
capture on nuclei and results in the reduction of the electron to
baryon ratio $Y_{e}$. The electron captures are strongly influenced
by the Gamow-Teller (GT+) transitions. In the late stage of the star
evolution, energies of the electrons are high enough to induce
transitions to the GT resonance. The importance of electron capture
for the presupernova collapse is also discussed in Ref.[5]. The
positron captures are of key importance in stellar core, especially
in high temperatures and low density locations. In such conditions,
a rather high concentration of positron can be reached from an
$e^{-} + e^{+}$ $\longleftrightarrow \gamma + \gamma$ equilibrium
which favor the e- e+ pairs. The competition (and perhaps
equilibrium) between positron captures on neutrons and electron
captures on protons is an important ingredient of the modeling of
Type-II supernovae. Recognizing the pivotal role played by capture
process, Fuller \emph{et al}. (referred as FFN) [6] calculated
systematically the electron and positron capture rates over a wide
range of temperature $(10^{7}\leq T(K)\leq 10^{11})$ and density
$(10\leq \rho Y_{e} \hspace{0.1in}(gcm^{-3})\leq 10^{11})$ for 226
nuclei with masses between A = 21 and 60. They stressed on the
importance of capture process to the GT resonance. The FFN rates
were then updated taking into account quenching of GT strength by an
overall factor of two by Aufderheide et al. [7]. The authors
stressed the need of a microscopic theory for calculation of
reliable rates vital for simulation codes of core collapse. Two
fully microscopic approaches, i.e., the shell model and
quasiparticle random phase approximation (QRPA), have been used
extensively for the large scale calculation of weak rates. In shell
model emphasis is more on interactions as compared to correlations
whereas QRPA puts more weight in correlations. Shell model
calculations are normally done taking a big core and some few
nucleons in the valence orbital. The QRPA calculations on the other
hand take all nucleons in the valence orbital and approximately none
in the core. Because of the large dimensionality of the space
involved for the pf-shell nuclei and beyond, Hamiltonian
diagonalization and calculation of beta decay strength is
computationally a formidable task. The Shell Model Monte Carlo
(SMMC) method was applied with relative success [e.g. 8, 9]. These
calculations, unfortunately, do not allow for detailed spectroscopy.
Secondly, the Monte Carlo path integral techniques are limited to
interactions that are free of the the "sign problem" and are still
computationally very intensive (i.e., requires supercomputer time).
The QRPA approach gives us the liberty of performing calculations in
a luxurious model space (as big as $7\hbar \omega $). Langanke and
collaborators [10] pointed out that QRPA is the method of choice for
dealing with heavy nuclei, and for predicting their half-lives, in
particular, based on the calculation of the GT strength function.
The QRPA method considers the residual correlations among nucleons
via one particle one hole (1p-1h) excitations in a large
multi-$\hbar \omega $  model spaces. An important extension of the
model in Ref. [11] includes the contribution of the configurations
more complex than 1p- 1h. Halbleib and Sorensen [12] for the first
time proposed and applied the pn-QRPA theory with separable GT (or
Fermi) interactions on spherical harmonic basis and later it was
extended to deformed nuclei [13, 14] using deformed single particle
basis. Nabi and Klapdor used the pn-QRPA theory to calculate the
stellar weak interaction rates over a wide range of temperature and
density scale for sd- [11] and fp/fpg-shell nuclei [15]. This work
is based on the pn-QRPA theory. We performed the evaluation of the
weak interaction rates and summed them over all parent and daughter
states to get the total rate. We considered a total of 30 excited
states in parent nucleus. The inclusion of a very large model space
of $7\hbar \omega$ in our model provides enough space to handle
excited states in parent and daughter nuclei (around 200) which
leads to satisfactory convergence of the electron capture rates (see
Eq.13). Transitions between these states play an important role in
the calculated weak rates. All previous compilations of weak
interaction rates either ignore transitions from parent excited
states due to complexity of the problem or apply the so-called
Brink's hypothesis when taking these excited states into
consideration. This hypothesis assumes that the Gamow-Teller
strength distribution on the excited states is same as for the
ground state, only shifted by the excitation energy of the state. We
do not use Brink's hypothesis to estimate the Gamow-Teller
transitions from parent excited states but rather we performed a
state-by-state evaluation of the weak interaction rates and summed
them over all parent and daughter sates to get the total weak rate.
This is the second major difference between this work and previous
calculations of electron capture rates. The result is an enhancement
of electron capture rates on $^{55}Co$ compared to the earlier
reported rates. Reliability of calculated rates is a key issue and
of decisive importance for many simulation codes. The reliability of
pn-QRPA model has already been established and discussed in detail
[11, 15, 16, 17]. There the authors compared the measured data of
thousands of nuclides with the pn-QRPA calculations and got good
comparison. In this paper, we calculate electron and positron
capture rates on $^{55}Co$ using the pn-QRPA theory. $^{55}Co$ is
abundant in the presupernova conditions, and as such is believed to
play a key role in the evolution of core collapse. Heger and
collaborators [18] identified $^{55}Co$ as the most important
nuclide for electron capture for massive stars ($25M_\odot$).
$^{55}Co$ is also considered among the top ten most important
electron capture nuclei during the presupernova evolution (see Table
25 of Aufderheide \emph{et al}. in Ref. [19]). In $\S$ 2 we discuss
the formalism for rate calculation. $\S$ 3 deals with calculation of
nuclear matrix elements. In $\S$ 4 we present and discuss the
results. Here we also compare our results with the previous
compilations. We finally summarize our discussions in $\S$ 5.

\section{Formalism}

The formalism used to
calculate weak rates at high temperatures and densities (relevant
to stellar environment) using the pn-QRPA theory is discussed in
this section. The following assumptions are made in the
calculation of weak rates.

(i) Only allowed GT and super-allowed Fermi transitions are
calculated. It is assumed that contributions from forbidden
transitions are relatively negligible.

(ii) The temperature is assumed high enough to ionize the atoms
completely. The electrons are not bound anymore to the nucleus and
obey the Fermi-Dirac distribution. At high temperatures (kT $>$ 1
MeV), positrons appear via electron-positron pair creation, and
positron follow the same energy distribution function as the
electrons.

(iii) The distortion of electron (positron) wavefunction due to
the coulomb interaction with a nucleus is represented by the Fermi
function in the phase space integrals.

(iv) Neutrinos and antineutrinos escape freely from the interior
of the star. Therefore, there are no (anti) neutrinos which block
the emission of these particles in the capture or decay processes.
Also, (anti)neutrino capture is not taken into
account.

 The Hamiltonian of our model is chosen as
\begin{equation} \label{GrindEQ__1_} H^{QRPA} =H^{sp} +V^{pair}
+V_{GT}^{ph} +V_{GT}^{pp}, \end{equation} Here $H^{sp} $ is the
single-particle Hamiltonian, $V^{pair} $ is the pairing force,
$V_{GT}^{ph} $ is the particle-hole (ph) Gamow- Teller force, and
$V_{GT}^{pp} $ is the particle particle (pp) Gamow-Teller force.
Wave functions and single particle energies are calculated in the
Nilsson model [20], which takes into account the nuclear
deformations. Pairing is treated in the BCS approximation. The
proton-neutron residual interactions occur in two different forms,
namely as particle-hole and particle-particle interaction. The
interactions are given separable form and are characterized by two
interaction constants $\chi$  and $\kappa$, respectively. The
selections of these two constants are done in an optimal fashion.
For details of the fine tuning of the Gamow-Teller strength
parameters, we refer to Ref. [21, 22]. In this work, we took the
values of $\chi = 0.2 MeV$ and $\kappa = 0.07 MeV$. Other
parameters required for the calculation of weak rates are the
Nilsson potential parameters, the deformation, the pairing gaps,
and the Q-value of the reaction. Nilsson-potential parameters were
taken from Ref. [23] and the Nilsson oscillator constant was
chosen $\hbar \omega=41A^{-1/3}(MeV)$, the same for protons and
neutrons. The calculated half-lives depend only weakly on the
values of the pairing gaps [24]. Thus, the traditional choice of
\[\Delta _{p} =\Delta _{n} =12/\sqrt{A} (MeV)\]
was applied in the present work.
The decay rates from the \emph{ith} state of the
parent to the \emph{jth} state of the daughter nucleus is given by
\begin{equation}
           \lambda _{ij} =\ln 2\frac{f_{ij} (T,\rho ,E_{f})}{(ft)_{ij} }
\end{equation}
\noindent where $(ft)_{ij}$ is related to the reduced transition
probability $B_{ij}$ of the nuclear transition by
\begin{equation}
             (ft)_{ij} =D/B_{ij}
\end{equation}
 \noindent D is a constant and is
\begin{equation}
 D=\frac{2\ln 2\hbar ^{7}\pi^{3}}{g_{v}^{2} m_{e}^{5}
c^{4} }
\end{equation}
\noindent and $B'_{ij}s$ are the sum of reduced transition
probabilities of the Fermi and GT transitions.
\begin{equation}
 B_{ij} =B(F)_{ij} +(g_{A}
/g_{V} )^{2} B(GT)_{ij}
\end{equation}

 We take the value of D
= 6295 s [22] and the ratio of the axial vector $(g_{A})$ to the
vector $(g_{v})$ coupling constant as - 1.254. Since then these
values have changed a little but did not lead to any significant
change in our rate calculations. The reduced transition
probabilities B(F) and B(GT) of the Fermi and GT transitions,
respectively, are given by
\begin{equation}
B(F)_{ij}=\frac{1}{2J_{i} +1} \left|\left\langle j\right. \left\|
\sum _{k}t_{\pm }^{k} \right\| \left. i\right\rangle \right|^{2}
\end{equation}
\begin{equation}
 B(GT)_{ij} =\frac{1}{2J_{i} +1}
\left|\left\langle j\right. \left\| \sum _{k}t_{\pm }^{k}
\overrightarrow{\sigma } ^{k} \right\| \left. i\right\rangle
\right|^{2}
\end{equation}

In Eq. (7),  $\overrightarrow{\sigma}(k)$ is the spin operator and
$t_{\pm }^{k} $ stands for the isospin raising and lowering
operator. The phase space integral $f_{ij}$ is an integral over
total energy and for electron and positron capture it is given by
\begin{equation}
f_{ij}=\int _{w_{1} }^{\infty }w\sqrt{w^{2} -1}  (w_{m} +w)^{2}
F(\pm Z,w)G_{\mp } dw.
\end{equation}

In Eq. (8), lower signs are for continuum positron capture and
upper signs are for electron capture. w is the total energy of the
electron including its rest mass, and $w_{l}$ is the total capture
threshold energy (rest + kinetic) for positron (or electron)
capture. $G_{-} (G_{+})$  is the electron (positron) distribution
function. These are the Fermi-Dirac distribution functions, with
\begin{equation}
 G_{-} =\left[\exp \left(\frac{E-E_{f} }{kT} \right)+1\right]^{-1}
\end{equation}
\begin{equation}
G_{+} =\left[\exp \left(\frac{E+2+E_{f} }{kT} \right)+1\right]^{-1}
\end{equation}
Here $E=(w - 1)$ is the kinetic energy of the electrons, $E_{f}$ is
the Fermi energy of the electrons, T is the temperature, and $k$ is
the Boltzmann constant. In Eq. 8, $F(Z,w)$ are the Fermi functions
and are calculated according to the procedure adopted by Gove and
Martin [25]. If the corresponding electron or positron emissions
total energy $(w_{m})$ is greater than -1, then $w_{l} = 1$, and if
less than or equal to 1, then $w_{l}=|w_{m}|$, where $w_{m}$ is the
total $\beta$ decay energy,
\begin{equation}
w_{m} =m_{p} -m_{d}+E_{i} -E_{j}
\end{equation}
\noindent where $m_{p}$ and $E_{i}$ are mass and excitation energies
of the parent nucleus, and $m_{d}$ and $E_{j}$ are mass and
excitation energies of the daughter nucleus, respectively. The
number density of electrons associated with protons and nuclei is
$\rho Y_{e}N_{A}$ ($\rho$ is the baryon density, $Y_{e}$ is electron
to baryon ratio, and $N_{A}$ is Avogadros number)
\begin{equation}
\rho Y_{e}=\frac{1}{\pi ^{2} N_{A} }(\frac{m_{e}c}{\hbar})^{3}\int
_{0}^{\infty }(G_{-}  -G_{+} )p^{2} dp
\end{equation}
 here
$p = (w^{2}-1)^{1/2}$ is the electron momentum and Eq. (12) has the units of
$mol \hspace{0.1in} cm^{-3}$. This equation is used for an iterative
calculation of Fermi energies for selected values of $Y_{e}$ and T.
There is a finite probability of occupation of parent excited states
in the stellar environment as result of the high temperature in the
interior of massive stars. Weak interactions then also have a finite
contribution from these excited states. The rate per unit time per
nucleus for any weak process is given by
\begin{equation}
\lambda =\sum _{ij}P_{i} \lambda _{ij},
\end{equation}
where $P_{i}$ is extracted using
\begin{equation}
P_{i} =\frac{(2J_{i} +1)\exp(-E_{i} /kT)}{\sum _{i=1} (2J_{i}
+1)\exp (-E_{i} /kT)}
\end{equation}
The summation in Eq. 13 is carried out over all initial and final
states until satisfactory convergence in our rate calculations is
achieved. The Fermi operator is independent of space and spin, and
as a result the Fermi strength is concentrated in a very narrow
resonance centered around the isobaric analogue state (IAS) for
the ground and excited states. The IAS is generated by operating
on the associated parent states with the isospin raising or
lowering operator:

 \[T_{\pm } =\sum _{i}t_{\pm } (i) ,\]
where the sum is over the nucleons. This operator commutes with
all parts of the nuclear Hamiltonian except for the coulomb part.
The superallowed Fermi transitions were assumed to be concentrated
in the IAS of the parent state. The Fermi matrix element depends
only on the nuclear isospin, T, and its projection \[T_{z}
=(Z-N)/2\]for the parent and daughter nucleus. The energy of the
IAS is calculated according to the prescription given in Ref.
[26], whereas the reduced transition probability is given by

\[B(F)=T(T+1)-T_{zi} T_{zf} ,\]
where $T_{zi}$  and $T_{zf}$  are the third components of the
isospin of initial and final analogue states, respectively.

\section{Calculation of Nuclear Matrix Elements}

 The RPA is formulated for excitations from the $J^{\pi}= 0$
ground state of an even-even nucleus. When the parent nucleus
has an odd nucleon, the ground state can be expressed as a
one-quasiparticle (q.p.) state, in which the odd q.p. occupies the
single-q.p. orbit of the smallest energy. Then two types of
transitions are possible. One is the phonon excitations, in which
the q.p. acts merely as a spectator. The other is transitions of the
q.p., where phonon correlations to the q.p. transitions in first
order perturbation are introduced [12]. The phonon-correlated
one-q.p. states are defined by

\begin{subequations}
\begin{eqnarray}
&& |p_{c} \rangle \, =\,
a_{p}^{+} {\left| - \right\rangle} \, + \, \sum _{n,\omega }a_{n}^{+}
A_{\omega }^{+} (\mu ){\left| - \right\rangle}
\times  \nonumber \\
&& |\left\langle - \right| \left[a_{n}^{+} A_{\omega }^{+}
(\mu )\right]^{+} H_{31} a_{p}^{+} {\left| - \right\rangle} E_{p} (n,\omega ), \\
&& |n_{c} \rangle \, = \, a_{n}^{+} {\left| - \right\rangle} \, +\,
\sum _{p,\omega }a_{p}^{+} A_{\omega }^{+} (-\mu ){\left| - \right\rangle}
\times  \nonumber \\
&& \left\langle - \right|
\left[a_{p}^{+} A_{\omega }^{+} (-\mu )\right]^{+} H_{31} a_{n}^{+}
{\left| - \right\rangle} E_{n} (p,\omega ) ,
\end{eqnarray}
\end{subequations}

\begin{equation}
E_{a} (b,\omega )\, =\,
\frac{1}{(\, \varepsilon _{a} \, -\,\varepsilon _{b} \, -\, \omega )}
\end{equation}
the first term of (15) is a proton (neutron) q.p. state and the
second term represents correlations of RPA phonons admixed by the
phonon-q.p. coupling Hamiltonian $H_{31}$, which is obtained from
the separable ph and pp forces by the Bogoliubov transformation
[27]. The sums run over all phonons and neutron (proton) q.p. states
which satisfy $m_{p}-m_{n} = \mu$, where $m_{p(n)}$ denotes the
third component of the angular momentum and $\pi_{p}.\pi_{n}= 1$.
Derivations of the q.p. transitions amplitudes for the correlated
states are given in Ref.[27] for a general force and a general mode
of charge-changing transitions. Also
\begin{equation}
{\left\langle n_{c}  \right|} t_{\pm } \sigma _{-\mu } {\left|
p_{c} \right\rangle} =(-1)^{\mu } {\left\langle p_{c}  \right|}
t_{\mp } \sigma _{\mu } {\left| n_{c}  \right\rangle}
\end{equation}
The excited states can be constructed as phonon-correlated
multi-quasiparticles states. The transition amplitudes between the
multi-quasiparticle states can then be reduced to those of
single-particle states. Low-lying states of an odd-proton
even-neutron nucleus ($^{55}Co$) can be constructed

(i) by exciting the odd proton from the ground state
(one-quasiparticles sates),

(ii) by excitation of paired proton (three proton states), or,

(iii) by excitation of a paired neutron (one-proton two-neutron
states). The multi-quasiparticle transitions can be reduced to
ones involving correlated (c) one-quasiparticle states:

\begin{eqnarray}
&&\left \langle p_{1}^{f} p_{2}^{f} p_{1c}^{f}
\left|t_{\pm } \sigma _{-\mu } \right|\right.
\left. p_{1}^{i} p_{2}^{i} n_{3c}^{i} \right\rangle = \nonumber \\
&& \delta (p_{1}^{f} ,p_{2}^{i} )\delta (p_{2}^{f},p_{3}^{i} )\,
\left\langle n_{1c}^{f} \left|t_{\pm } \sigma _{-\mu } \right|\right.
\left. p_{1c}^{i} \right\rangle  \nonumber \\
&& -\delta (p_{1}^{f} ,p_{1}^{i} )\delta (p_{2}^{f} ,p_{3}^{i} )\,
\left\langle n_{1c}^{f} \left|t_{\pm } \sigma _{-\mu } \right|\right. \left.
p_{2c}^{i} \right\rangle  \nonumber \\
&& +\delta (p_{1}^{f} ,p_{1}^{i} )\delta (p_{2}^{f} ,p_{2}^{i} )\left\langle n_{1c}^{f} \left|t_{\pm }
\sigma _{-\mu } \right|\right. \left. p_{3c}^{i} \right\rangle
\end{eqnarray}

\begin{eqnarray}
&& {\left\langle p_{1}^{f} p_{2}^{f} n_{1c}^{f} \left|t_{\pm }
\sigma _{\mu } \right|\right. \left. p_{1}^{i} n_{1}^{i} n_{2c}^{i}
 \right\rangle =}\nonumber \\
&& \delta (n_{1}^{f} ,n_{2}^{i} )\left[\delta (p_{1}^{f},
p_{1}^{i} )\left\langle p_{2c}^{f} \left|t_{\pm } \sigma _{\mu }
\right|\right. \left. n_{1c}^{i} \right\rangle \right. \nonumber  \\
&& {\left. \, \, \, -\delta (p_{2}^{f} ,p_{1}^{i} )
\left\langle p_{1c}^{f} \left|t_{\pm } \sigma _{\mu } \right|\right.
\left. n_{1c}^{i} \right\rangle \right]-\delta (n_{1}^{f} ,n_{1}^{i} )}
\nonumber \\
&&{\times \left[\delta (p_{1}^{f} ,p_{1}^{i} )\left\langle p_{2c}^{f}
\left|t_{\pm } \sigma _{\mu } \right|\right. \left. n_{2c}^{i}
\right\rangle \right. -\delta (p_{2}^{f} ,p_{1}^{i} )} \nonumber \\
&&{ \left. \times \left\langle p_{1c}^{f} \left|t_{\pm }
\sigma _{\mu } \right|\right. \left. n_{2c}^{i}
\right\rangle \right]}
\end{eqnarray}

\begin{eqnarray}
&&{\left\langle n_{1}^{f} n_{2}^{f} n_{3c}^{f} \left|t_{\pm }
\sigma _{-\mu } \right|\right. \left. p_{1}^{i} n_{1}^{i} n_{2c}^{i}
\right\rangle =} \nonumber \\
&&{ \delta (n_{2}^{f} ,n_{1}^{i} )\delta (n_{3}^{f} ,n_{2}^{i} )\,
\left\langle n_{1c}^{f} \left|t_{\pm } \sigma _{-\mu } \right|\right.
\left. p_{1c}^{i} \right\rangle } \nonumber \\
&&{ -\delta (n_{1}^{f} ,n_{1}^{i} )\delta (n_{3}^{f} ,n_{2}^{i} )\,
\left\langle n_{2c}^{f} \left|t_{\pm } \sigma _{-\mu } \right|\right.
\left. p_{1c}^{i} \right\rangle } \nonumber  \\
&&{+\delta (n_{1}^{f} ,n_{1}^{i} )\delta (n_{2}^{f} ,n_{2}^{i} )\left\langle
 n_{3c}^{f} \left|t_{\pm } \sigma _{-\mu } \right|\right.
\left. p_{1c}^{i} \right\rangle .}
\end{eqnarray}

For odd-neutron even-proton nucleus ($^{55}Fe$, $^{55}Ni$) the
excited states can be constructed

(i) by lifting the odd neutron from the ground state to excited
states (one  quasiparticle state),

(ii) By excitation of a paired neutron (three neutron states), or,

(iii) by the excitation of a paired proton (one-neutron two-proton
states). Once again the multi-qp states are reduced to ones
involving only correlated (c) one-qp states:

\begin{eqnarray}
&& {\left\langle p_{1}^{f} n_{1}^{f} n_{2c}^{f} \left|t_{\pm }
\sigma _{\mu } \right|\right. \left. n_{1}^{i} n_{2}^{i} n_{3c}^{i}
\right\rangle =} \nonumber \\
&& {  \delta (n_{1}^{f} ,n_{2}^{i} )\delta (n_{2}^{f} ,n_{3}^{i} )
\left\langle p_{1c}^{f} \left|t_{\pm } \sigma _{\mu } \right|\right.
\left. n_{1c}^{i} \right\rangle } \nonumber \\
&& { -\delta (n_{1}^{f} ,n_{1}^{i} )\delta (n_{2}^{f} ,n_{3}^{i} )
\left\langle p_{1c}^{f} \left|t_{\pm } \sigma _{\mu } \right|\right.
\left. n_{2c}^{i} \right\rangle } \nonumber \\
&& { +\delta (n_{1}^{f} ,n_{1}^{i} )\delta (n_{2}^{f} ,n_{2}^{i} )\left
\langle p_{1c}^{f} \left|t_{\pm } \sigma _{\mu }
\right|\right. \left. n_{3c}^{i} \right\rangle }
\end{eqnarray}

\begin{eqnarray}
&&{\left\langle p_{1}^{f} n_{1}^{f} n_{2c}^{f} \left|t_{\pm } \sigma _{-\mu }
\right|\right. \left. p_{1}^{i} p_{2}^{i} n_{1c}^{i} \right\rangle =}\nonumber \\
&&{ \delta (p_{1}^{f} ,p_{2}^{i} )\left[\delta (n_{1}^{f} ,n_{1}^{i} )\right.
 \left\langle n_{2c}^{f} \left|t_{\pm } \sigma _{-\mu } \right|\right.
\left. p_{1c}^{i} \right\rangle } \nonumber \\
&&{ -\delta (n_{2}^{f} ,n_{1}^{i} )\left\langle n_{1c}^{f} \left|t_{\pm }
\sigma _{-\mu } \right|\right. \left. p_{1c}^{i} \right\rangle
-\delta (p_{1}^{f} ,p_{1}^{i} )} \nonumber \\
&&{ \times \left[\delta (n_{1}^{f} ,n_{1}^{i} )\right.
\left\langle n_{2c}^{f} \left|t_{\pm } \sigma _{-\mu } \right|\right. \left.
p_{2c}^{i} \right\rangle -\delta (n_{2}^{f} ,n_{1}^{i} )} \nonumber \\
&&{ \left. \times \left\langle n_{1c}^{f} \left|t_{\pm }
\sigma _{-\mu } \right|\right. \left. p_{2c}^{i}
\right\rangle \right].}
\end{eqnarray}

\begin{eqnarray}
&& {\left\langle p_{1}^{f} p_{1}^{f} p_{3c}^{f} \left|t_{\pm } \sigma _{\mu }
 \right|\right. \left. p_{1}^{i} p_{2}^{i} n_{1c}^{i} \right\rangle =}
\nonumber \\
&&{ \delta (p_{2}^{f} ,p_{1}^{i} )\delta (p_{3}^{f} ,p_{2}^{i} )
\left\langle p_{1c}^{f} \left|t_{\pm } \sigma _{\mu } \right|\right.
\left. n_{1c}^{i} \right\rangle } \nonumber \\
&&{ -\delta (p_{1}^{f} ,p_{1}^{i} )\delta (p_{3}^{f} ,p_{2}^{i} )
\left\langle p_{2c}^{f} \left|t_{\pm } \sigma _{\mu }
\right|\right. \left. n_{1c}^{i} \right\rangle } \nonumber \\
&&{ +\delta (p_{1}^{f} ,p_{1}^{i} )\delta (p_{2}^{f} ,p_{2}^{i} )
\left\langle p_{3c}^{f} \left|t_{\pm } \sigma _{\mu }
\right|\right. \left. n_{1c}^{i} \right\rangle .}
\end{eqnarray}

GT transitions of phonon excitations for every excited state were
also taken into account. We also assumed that the quasiparticles in
the parent nucleus remained in the same quasiparticle orbits.
Further details can be found in Ref. [22].

\section{Results and Discussions }

\begin{figure}[htb]
\includegraphics[width=7cm]{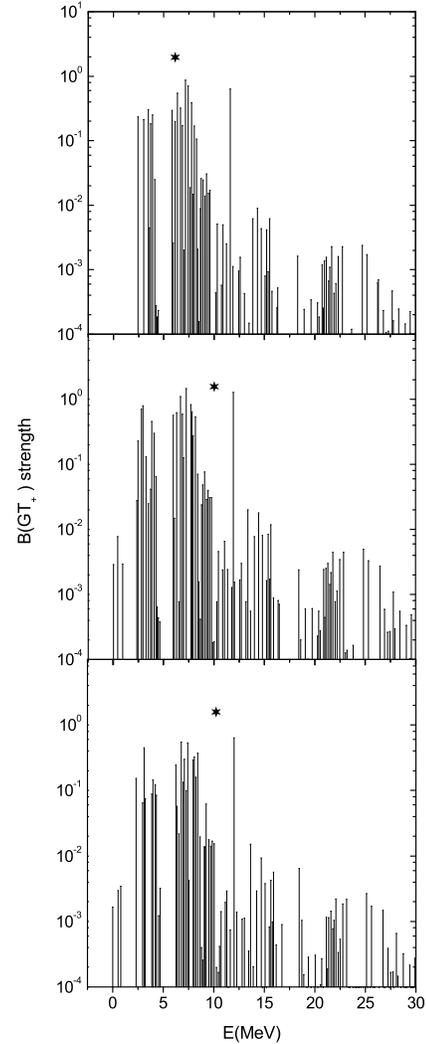}
\caption{
Gamow-Teller $(GT_{+})$ strength distribution for electron
captures on $^{55}$Co. The top panel shows GT strength for ground
state, whereas middle and bottom panels show GT strength for first
and second excited states, respectively. The GT centroids in Ref.
[28] are indicated by asterisks for the respective states. The
energy scale is the excitation energies in daughter.}
\end{figure}
\begin{figure}[htb]
\includegraphics[width=7cm]{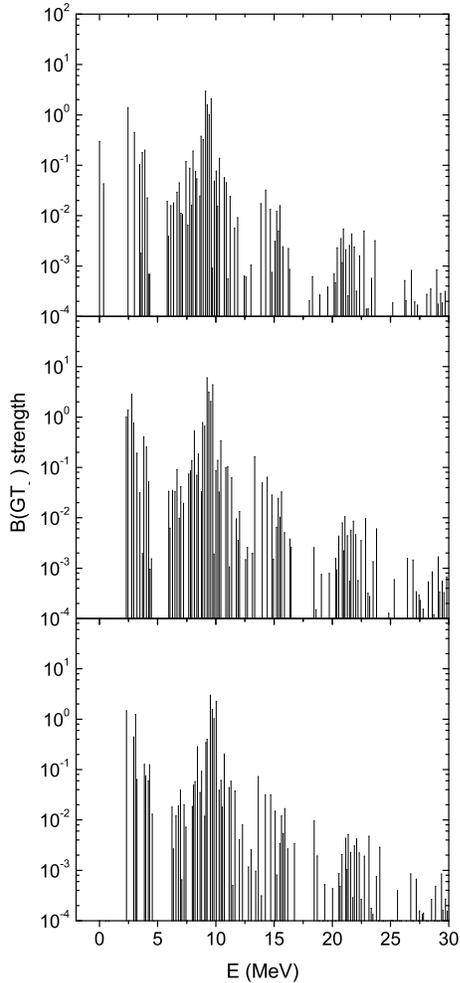}
\caption{
 Gamow-Teller $(GT_{-})$ strength distribution for positron
captures on $^{55}$Co. From top to bottom, the panels show GT-
strength for ground, first and second excited states, respectively.
Energy scale refers to excitation energies in daughter.}
\end{figure}
\begin{figure}[htb]
\includegraphics[width=7cm]{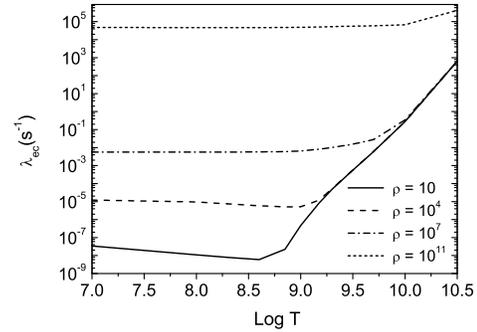}
\caption{
Electron capture rates on $^{55}$Co as function of
temperature for different selected densities. Densities are in units
of $g cm^{-3}$. Temperatures are measured in $K$.}
\end{figure}
\begin{figure}[htb]
\includegraphics[width=8.4cm]{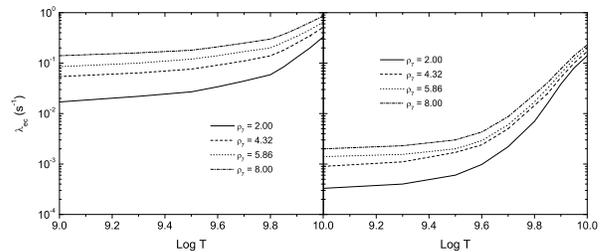}
\caption{
 Electron capture rates on $^{55}$Co as
function of temperature for different densities (left panel). The
right panel shows the results of Ref. [28] for the corresponding
temperatures and densities. For units see text.}
\end{figure}
\begin{figure}[htb]
\includegraphics[width=7cm]{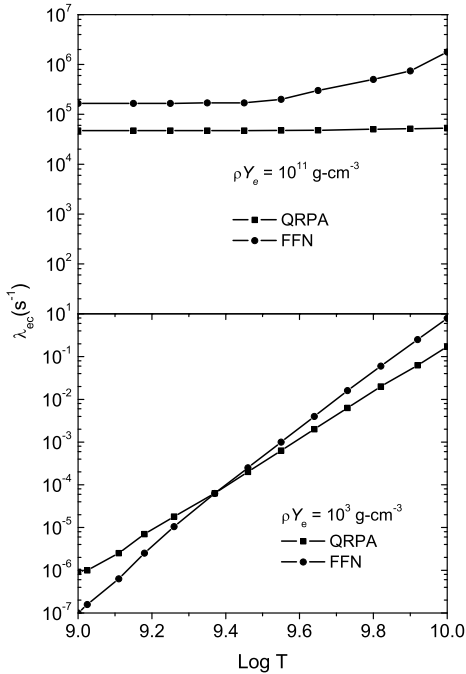}
\caption{
Comparison of our QRPA electron capture rates
on $^{55}$Co with FFN rates [6] for densities $\rho Y_{e} = 10^{3} g
cm^{-3}$ (lower panel) and $\rho Y_{e} = 10^{11} g cm^{-3}$ (upper
panel).}
\end{figure}
\begin{figure}[htb]
\includegraphics[width=7cm]{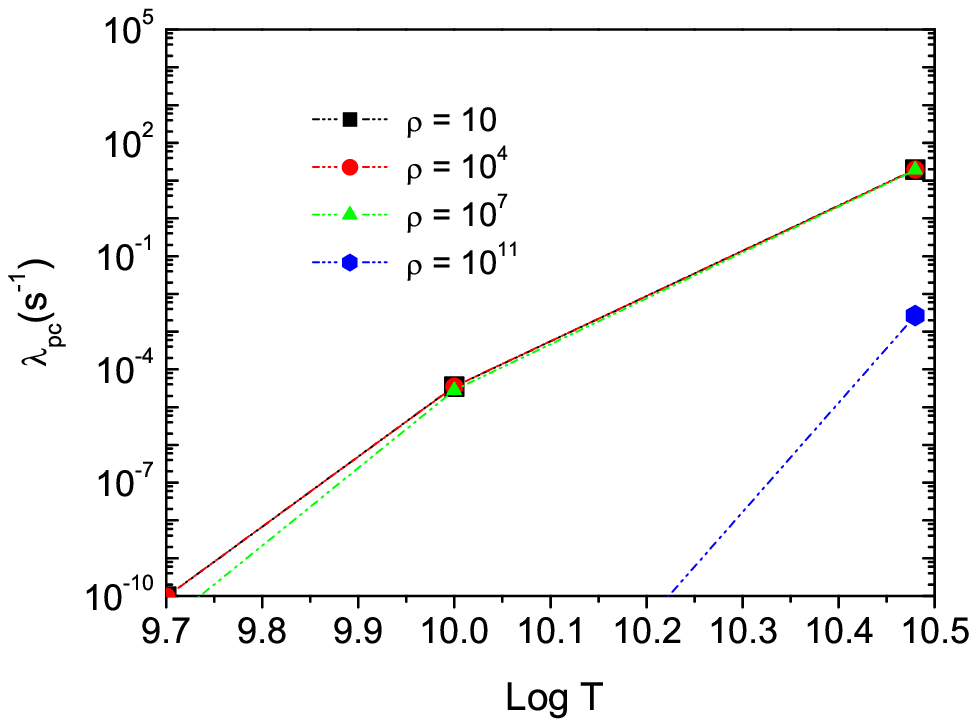}
\caption{
 Positron captures on $^{55}$Co as function of
temperatures and densities. Densities in inset are in units of $g
cm^{-3}$. Temperatures are measured in $K$.}
\end{figure}

 For the parent nuclide we
considered a maximum of 30 states. States still higher in excitation
energy were not considered as their occupation probability was not
high enough for the temperature and density scales chosen for this
phase of core collapse. For each parent state we considered around
200 states of daughter. GT strength of each contributing state was
taken into account. GT transitions are the dominant excitation mode
for electron captures during the presupernova evolution. The B(GT)
strength distributions for ground and two excited states at 2.2 MeV,
and 2.6 MeV are shown in Figs.~1 and 2, respectively. We note that
the GT strength is fragmented over
many daughter states. For
electron capture, the GT centroid resides in the energy range
7.1 and 7.4 MeV in the daughter $^{55}Fe$ and more or less in the energy
range 6.7 - 7.5 MeV for both the excited states. We get a good
energy resolution in our spectra which we attribute to the large
model space. The corresponding values for the ground and the two
excited states using shell model [28] are 6 MeV and 9 and 10 MeV,
respectively, and are also shown by asterisk in the figure. We
clearly see from Fig.~1 that our centroid is shifted to much lower
energies for the excited states. Our code also calculated GT
transitions for the states at 0.3 MeV and 0.4 MeV which are close
lying states with the ground state. The centroids for these states
are in the range of 6.6 and 7.4 MeV and 7.2 and 8.1 MeV, respectively.
Transitions from these low-lying states contributed to the
enhancement of our electron capture rates. For positron captures,
the GT centroid resides at 9.1 MeV for the ground state. The
corresponding values of the centroids for the 2.2 MeV and 2.6 MeV
excited states are 9.2 MeV and 9.6 MeV, respectively. In some
situations, the total GT strength is more important than the GT
centroid. We worked out the total GT strength for the electron
capture on $^{55}Co$ to be 7.4 and 17.9 for positron capture. The
variation of electron capture rates for $^{55}Co$ with temperatures
and densities is shown in Fig.~3. The temperature scale log T
measures the temperature in K and the density shown in the inset has
units of $gcm^{-3}$. In low density $(10-10^{4} gcm^{-3})$ regions
of the star, the electron capture rates on $^{55}Co$ nuclide
decreases as temperature of the stellar core increases. This trend
continues until log T is in the vicinity of 8.6. Beyond this
temperature electron capture rates shoots up. In low temperature
regions (log T = 7.0), when the stellar core shift from densities
$\rho= 10^{7} gcm^{-3}$ to $10^{11} gcm^{-3}$, the electron capture
rates are enhanced by as much as 7 order of magnitude. For high
densities $(\rho = 10^{11} gcm^{-3})$, the electron capture rates
remain constant until around log T = 10. Above this temperature
enhancement of the electron capture rates take place. The electron
capture rates for different densities grid and temperatures can be
seen from Table~I.
\begin{table}
\caption{ Electron Capture Rates $^{55}Co \longrightarrow ^{55}Fe$}
\begin{tabular}{cccccc}\hline\\
$\rho(gcm^{-3})$/logT&  7  & 8.6  & 9.2 &  9.7 &  10.5\\\hline
  10      & 3.4(-8) & 6.0(-9) & 8.3(-6) & 6.0(-3) & 6.3(2)\\
 $10^{4}$ & 1.2(-5) & 5.8(-6) & 1.2(-5) & 6.1(-3) & 6.3(2)\\
 $10^{7}$ & 5.7(-3) & 5.7(-3) & 8.2(-3) & 2.8(-2) & 6.4(2)\\
 $10^{11}$& 4.7 (4) & 4.7(4)  & 5.1(4)  & 5.9(4)  & 4.2(5)\\\hline
\end{tabular}

Electron capture rates for $^{55}Co$ as function of temperature
and density. \\
The values in the parenthesis represent the power
of 10. The units of each rates are $sec^{-1}$.
\end{table}
The units of each rate are $sec^{-1}$. In Fig.~4
we compare our electron capture rates with that of Ref. [28]. Here
$\rho_{7}$ measure the densities in $10^{7} gcm^{-3}$. We note that
our rates are two orders of magnitude faster at low temperature as
compared to Ref. [28]. Due to the complexity of the spectroscopy
involved, authors in Ref. [28] had to switch to approximations like
back resonances (the GT back resonance are states reached by the
strong GT transitions in the electron capture process built on
ground and excited states [6, 18]) and Brink's hypothesis. The
enhancement of our electron capture rates at presupernova
temperatures is due to large GT+ transitions from the low-lying
states of the parent nucleus. These states have finite probability
of occupation at presupernova temperatures. We do not assume the
Brink's hypothesis. Our results show that the Brink's hypothesis is
a first order approximation and much of the times transitions from
excited states are many orders of magnitude higher than those from
the ground state [15]. Low-lying transitions are quite important at
low temperatures and densities and supplement the electron capture
rate from the GT resonance if the Q value only allows capture of
high energy electrons from the tail of the Fermi- Dirac
distribution. We see from the GT distribution (Fig.~1) that the GT
centroid of the QRPA is very close to the shell model centroid, but
our centroids for the excited states are at low energy in daughter
as compared to shell model centroids. Contribution to rates from
these states is many orders of magnitude larger than the ground
state leading to an overall enhancement of our rates at low
temperatures. At supernova temperatures the difference between the
two calculations decreases. At higher temperatures and densities the
energy of the electron is large compared to Q value for transitions
to GT centroid. In such conditions the capture rates are no more
dependent on the energy of the GT distribution but rather depend on
the total GT strength [29]. The total GT strength of the QRPA, 7.4,
is less than the shell model, 8.7. Our rates are still around four
times faster than shell model ones at higher densities and
temperatures ([see also Ref. [30]). We calculated a $^{55}Co$
halflife of 1108 sec. (18.74 hours), which is in good agreement with
the experimental value of 17.53 hours [31]. FFN calculated stellar
electron and positron capture rates for 226 nuclei with masses
between A = 21 and 60. Measured nuclear level information and matrix
elements available at that time were used and unmeasured matrix
elements for allowed transitions were assigned an average value of
log ft = 5. To complete the FFN rate estimate, the Gamow-Teller
contribution to the rate was parameterized on the basis of the
independent particle model and supplemented by a contribution
simulating low-lying transitions. Fig.~5 shows the comparison of our
rates with the FFN rates [6] for densities $\rho Y_{e} = 10^{3}
gcm^{-3}$ and $\rho Y_{e} = 10^{11} gcm^{-3}$, respectively. For low
densities $(\rho Y_{e} = 10^{3}gcm^{-3})$ and temperatures our
electron capture rates are enhanced by one order of magnitude than
the FFN rates. Both rates increase with increasing temperature. We
are in good agreement with the FFN when temperature of the stellar
core is around log T = 9.4. Above this temperature, the FFN rates
are enhanced than our rates. The enhancement in FFN rates become
more pronounced at densities $\rho Y_{e} =10^{11}gcm^{-3}$. The main
reason for this enhancement is the placement of the GT centroid at
too low excitation energies by FFN as also pointed by authors in
Ref. [28]. The competition between positron captures on neutrons and
electron captures on protons is thought to play a crucial role in
modeling of Type-II supernovae [19]. The positron captures are also
of importance in stellar core having low density locations and
enough high temperatures. The continuum positron capture and
electron capture are characteristic of stellar plasma. Our positron
capture rates are shown in Fig.~6. We note that around presupernova
temperatures the positron capture rates are very slow as compared to
the electron capture rates. We assumed in our calculations that
positrons appear via electron-positron pair creation only when the
stellar temperature exceeds 1 MeV. When the temperature of the
stellar core increases further the positron capture rates shoot up.
In high temperature regions of the stellar core, capture rates are
more sensitive to the total GT strength distribution. We computed
the total GT strength around 17.9. This results in high capture
rates of positrons at low densities and high temperatures regions of
stars. The positron capture rates decreases with increasing
densities, in contrast to the electron capture rates which increase
as density increases. As temperature rises, more and more positrons
are created leading in turn to higher capture rates. Table II shows
our calculations of positron capture rates (in units of $sec^{-1}$)
at selected temperatures and densities.
\begin{table}
\caption{Positron Capture Rates
$^{55}Co \longrightarrow ^{55}Ni$}
\begin{tabular}{cccccc}\hline\\
$\rho(gcm^{-3})$/logT&  7  & 8.6 & 9.2 &  9.7 &  10.5\\\hline
  10      & 0.0 & 0.0 & 3.2(-33) & 9.6 (-11) & 1.9 (1)\\
 $10^{4}$ & 0.0 & 0.0 & 2.2(-33) & 9.5 (-11) & 1.9 (1)\\
 $10^{7}$ & 0.0 & 0.0 & 3.7(-37) & 1.9 (-11) & 1.9 (1)\\
 $10^{11}$& 0.0 & 0.0 & 0.0      & 7.6 (-35) & 2.7 (-3)\\\hline
\end{tabular}

Positron capture rates for $^{55}Co$ as function of temperature
and density. \\
The values in the parenthesis represent the power
of 10. The units of each rates are $sec^{-1}$.
\end{table}

 \section{Summary}

 Electron and positron capture rates on
$^{55}Co$ was calculated microscopically using the pn-QRPA theory,
which has been used extensively for calculation of terrestrial weak
rates with success. The pn-QRPA theory was used to calculate weak
rates in stellar environment. This theory also gave us the liberty
of using a large model space of $7\hbar\omega$. A total of 30 parent and around
200 daughter excited states for each parent state were considered in
our calculations. For each pair of calculated parent and daughter
states, the B(GT) strength was calculated in a microscopic fashion.
$^{55}Co$ is considered a strong candidate among the other Fe peak
nuclei that play a dominant role in electron capturing and hence in
the core collapse of a star. At presupernova the dynamics of the
star is very complex and large numbers of nuclear excited states are
involved. The pn-QRPA is a judicious choice for handling these large
numbers of excited states in heavy nuclei in the presupernova
conditions of the stellar core. Our results point to a much more
enhanced capture rate for $^{55}Co$ as compared to the reported
shell model rates and can have a significant astrophysical impact on
the core collapse simulations. The reduced capture rates for
$^{55}Co$ in the outer layers of the core from the previous
compilations resulted in slowing the collapse and posed a large
shock radius to deal with [2]. What impact our enhanced rates may
have on the core collapse simulations? According to Aufderheide
\emph{et al}. [19], the rate of change of lepton-to-baryon ratio (
$\dot{\psi{_e}}$ ) changes by about 50\% alone due to electron
capture on $^{55}Co$. Our results might point towards favoring a
prompt explosion. One cannot conclude just on the basis of one kind
of nucleus about the dynamics of explosion(prompt or delayed). We
recall that it is the rate and abundance of particular specie of
nucleus that prioritizes the importance of that particular nucleus
in controlling the dynamics of late stages of stellar evolution.


\begin{thebibliography}{99}

\bibitem{[1]} H.A. Bethe, G.E. Brown, J. Applegate, and J.M. Lattimer, Nucl.
Phys. A \textbf{324}, 487 (1979).

\bibitem{[2]} W.R. Hix, O.E. Messer, A.
Mezzacappa, M. Liebendrfer, J. Sampio, K. Langanke, D.J. Dean, and G.
Mart\'\i nez-Pinedo, Phys. Rev. Lett. \textbf{91}, 201102 (2003).

\bibitem{[3]}  F.K. Sutaria, A. Ray, J.A. Sheikh, and P. Ring, Astron.
Astrophys. \textbf{349}, 135 (1999).

\bibitem{[4]} A. Heger, N. Langer,
S.E. Woosley, Ap. J. \textbf{528}, 368 (2000).

 \bibitem{[5]} H.A. Bethe,
Rev. Mod. Phys. \textbf{62}, 801 (1990).

\bibitem{[6]} G. M. Fuller, W. A.
Fowler, and M. J. Newman, ApJS \textbf{42}, 447 (1980); \textbf{48},
279 (1982); ApJ \textbf{252}, 715 (1982).

\bibitem{[7]} M.B. Aufderheide,
G.E. Brown, T.T.S. Kuo, D.B. Stout, and P. Vogel, ApJ\textbf{362}, 241
(1990).

\bibitem{[8]} C.W. Johnson, S.E. Koonin, G.H. Lang, and  W.E. Ormand,
Phys. Rev. Lett. \textbf{69}, 3157 (1992).

\bibitem{[9]}  S.E. Koonin,
D.J. Dean, and K. Langange, Phys. Rep. \textbf{278}, 1 (1996).

 \bibitem{[10]}  K. Langanke, G. Martinez-Pinedo,
Rev. Mod. Phys. \textbf{75}, 819 (2003).

 \bibitem{[11]}  J.-U. Nabi, H. V.
Klapdor- Kleingrothaus, Atomic Data and Nuclear Data Tables
\textbf{71}, 149 (1999).

\bibitem{[12]}  J. A. Halbleib, R. A. Sorensen,
Nucl. Phys. A \textbf{98}, 542 (1967).

 \bibitem{[13]}  J. Krumlinde, P. Muller, Nucl. Phys. A \textbf{417},
419 (1984).

\bibitem{[14]} P. Muller, J. Randrup, Nucl. Phys.
A \textbf{514}, 1 (1990).

\bibitem{[15]}  J.-U. Nabi, H.V. Klapdor-
Kleingrothaus, Atomic Data and Nuclear Data Tables \textbf{88},
237 (2004).

\bibitem{[16]}  J.-U. Nabi, H. V. Klapdor- Kleingrothaus, Eur.
Phys. J. A \textbf{5}, 337 (1999).

\bibitem{[17]} J.-U. Nabi, Ph.D. Thesis,
Heidelberg University Germany, 1999.

 \bibitem{[18]}  A. Heger, K. Langanke, G.
Martínez-Pinedo, and S.E. Woosley, Phys. Rev. Lett. \textbf{86}, 1678
(2001).

\bibitem{[19]}  M.B. Aufderheide, I. Fushiki, S.E. Woosley, E.
Stanford, and D.H. Hartmann, Astrophys. J. Suppl. \textbf{91}, 389
(1994).

 \bibitem{[20]}
 S.G. Nilsson, Mat. Fys. Medd. Dan. Vid. Selsk \textbf{29}, No.16 (1955).

\bibitem{[21]}  A. Staudt, E. Bender, K. Muto, and H.V. Klapdor-Kleingrothaus,
Atomic Data and Nuclear Data Tables \textbf{44}, 79 (1990).

\bibitem{[22]}
M. Hirsch, A. Staudt, K. Muto, and H.V. Klapdor-Kleingrothaus, Atomic
Data and Nuclear Data Tables \textbf{53}, 165 (1993).

\bibitem{[23]}  I. Ragnarsson, R.K. Sheline, Phys. Scr. \textbf{29}, 385 (1984).

\bibitem{[24]} M. Hirsch, A. Staudt, K. Muto, and  H.V. Klapdor-Kleingrothaus,
Nucl. Phys. A \textbf{535}, 62 (1991).

\bibitem{[25]} N.B. Gove, M.J.
Martin, Atomic Data and Nuclear Data Tables \textbf{10}, 205
(1971).

\bibitem{[26]} K. Grotz, H.V. Klapdor, The Weak Interaction in
Nuclear, Particle and Astrophysics, Adam Hilger, (IOP Publishing,
Bristol, Philadelphia, New York, 1990).

\bibitem{[27]} K. Muto, E. Bender,
and H.V. Klapdor- Kleingrothaus, Z. Phys. A \textbf{334}, 187 (1989).

\bibitem{[28]} K. Langanke, G. Martinez-Pinedo, Phys. Lett. B \textbf{436},
19 (1998).

\bibitem{[29]} K. Langanke, G. Martinez-Pinedo, Nucl. Phys. A
\textbf{673}, 481 (2000).

\bibitem{[30]}  J.-U. Nabi, M.-U. Rahman, Phys.
Lett. B \textbf{612}, 190 (2005).

\bibitem{[31]} G. Audi, O. Bersillon, J.
Blachot, and A.H. Wapstra, Nucl. Phys. A \textbf{729}, 3 (2003).





\end{thebibliography}
\end{document}